%% file: paper.tex
\documentclass[twocolumn,showpacs]{revtex4-1}

\date{}

\usepackage{graphicx}% Include figure files
\usepackage{dcolumn}% Align table columns on decimal point
\usepackage{bm}% bold math
\usepackage{enumitem}
\usepackage{gensymb}
\usepackage{amsmath}
\usepackage{amssymb}
\usepackage{dcolumn}
\usepackage{color}
\begin{document}

\title{\bf Boosting the accuracy and speed of quantum Monte Carlo: size-consistency and time-step }

\author{Andrea Zen$^{1,2,3}$}
\author{Sandro Sorella$^4$}
\author{Michael J. Gillan$^{1,2,3}$}
\author{Angelos Michaelides$^{1,2,3}$}
\author{Dario Alf\`{e}$^{1,2,3,5}$}%
\email{d.alfe@ucl.ac.uk}

\affiliation{
 $^1$ London Centre for Nanotechnology, Gordon St., London WC1H 0AH, UK
 $^2$ Thomas Young Centre, University College London, London WC1H 0AH, UK
 $^3$ Dept. of Physics and Astronomy, University College London, London WC1E 6BT, UK
 $^4$ International School for Advanced Studies (SISSA), Via Beirut 2-4, 34014 Trieste, Italy and INFM Democritos National Simulation Center, Trieste, Italy
 $^5$ Dept. of Earth Sciences, University College London, London WC1E 6BT, UK
  }

             %  but any date may be explicitly specified}

\begin{abstract}
Diffusion Monte Carlo (DMC) simulations for fermions are becoming the standard 
for providing high quality reference data in systems that are too large to be investigated via quantum chemical approaches.
DMC with the fixed-node approximation relies on modifications of the Green function to avoid singularities near the nodal surface of the trial wavefunction. 
Here we show that these modifications affect the DMC energies in a way that is not size-consistent, resulting in large time-step errors. 
Building on the modifications of Umrigar {\em et al.} and  DePasquale {\em et al.} 
we propose a  simple Green function modification that restores size-consistency to large values of the time-step, which substantially reduces time-step errors. 
The new algorithm also yields remarkable speedups of up to two orders of magnitude in the calculation of molecule-molecule binding energies and crystal cohesive energies, 
thus extending the horizons of what is possible with DMC.
\end{abstract}

\maketitle

% motivation (Angelos)

The determination of accurate reference energetics for solids  is  one of the grand challenges of materials modelling. Reliable reference data is needed to make accurate predictions about any number of phenomena, such as phase stability, adsorption on surfaces and crystal polymorph prediction. Very often density functional theory (DFT) provides sufficient accuracy for this and as such has been immensely successful in furthering our understanding of materials~\cite{Hafner:2011fg,Neugebauer:2013dw}. %,fancy DFT genome papers}. 
However, there are many materials and materials related problems for which DFT does not deliver the desired accuracy~\cite{challengesDFT:2012}. For such problems explicitly correlated wave-function based approaches are needed, such as the approaches of quantum chemistry, quantum Monte Carlo (QMC), and combinations thereof \cite{foulkes01,reviewQC,CCSDT:Rev2007,DMRG:2002,FCIQMC:JCP2009,FCIQMC:Nat2013,AFQMC:Zhang2003,LRDMC:prl2005,casula10,RPA_Kresse:PRL2009,RPA_Schimka:NatMatt2010,RPA:JMS2012}. In practice for condensed phase systems with more than a handful of atoms in the unit cell QMC remains the only feasible reference method, partly because of its favorable scaling with system size and the fact that it can be used efficiently on massively parallel supercomputers. 
Indeed QMC, mostly within the diffusion Monte Carlo (DMC) approach, is increasingly used to provide benchmark data for solids and to tackle interesting materials science problems that have been beyond the reach of DFT~\cite{Ice:prl2011,Morales:bulkwat:2014,Cox:2014,Benali:2014,AlHamdani:hBN:2015,gillan15,AFQMC:CoGraf,Morales:perspective2014,Mazzola:nat2014,Mazzola:prl2015,Zen:liqwat2015,Chen:jcp2014,Wagner:2013,Wagner:prb2014}.
%some examples of fancy QMC applications, e.g. our ice and water adsorption papers and some others from groups? of likely referees
DMC is also proving increasingly useful in exposing and helping to explain problems with DFT and as such in helping to further the development of DFT. 
%However, because of the enormous cost of QMC in practical applications of QMC trade off is made 

DMC is in principle an exact technique to solve the imaginary time dependent Schr\"odinger equation. The discretization of time in practical implementations introduces a time-step ($\tau$) error, the computational cost of which is proportional to $1/\tau$.
%
%, namely to project out the ground state wave-function $\phi$ of an Hamiltonian $H$. 
%In actual DMC simulations a time-step $\tau$ is chosen, that is a trade-off between accuracy (exact results come for $\tau  \to 0$) and the computational cost ($\propto 1/\tau$).
% The purpose of this letter is to highlight a general problem that has recently come to light in the evaluation of binding energies of molecules: in current common implementations the DMC method appears to be non size-consistent for finite $\tau$, as pointed out by Gillan {\em et al.}~\cite{gillan15} in the context of the binding energies of CH$_4$ molecules in clusters of H$_2$O of various sizes. 
Recently Gillan {\em et al.}~\cite{gillan15} showed that for CH$_4$-H$_2$O clusters current implementations of DMC appear to be non size-consistent, i.e. the total energy of a system of $M$ non-interacting molecules is not proportional to $M$. 
%at finite $\tau$. 
Here we show that this is a general problem, we identify its source, and propose a simple modification that solves it. 
Moreover, we observe that the time-step error in binding energy evaluations is mostly due to this size-consistency issue. 
Our proposed method also leads to remarkable speedups, by significantly increasing the accuracy of large $\tau$ DMC evaluations
\footnote{We note that other QMC approaches, such as the variational Monte Carlo (VMC) or the lattice regularized diffusion Monte Carlo (LRDMC)~\cite{LRDMC:prl2005}
do not suffer from these problems. This has been shown in \cite{casula10}, 
where the effect of the cutoff in the local energy on the size-consistency issue was carefully considered also for the latter method. In this paper, however, we are concerned with the much more widely used  DMC.}.

A review of DMC can be found elsewhere~\cite{foulkes01,umrigar93}, and is summarized in Appendix~\ref{app:revDMC}. 
To understand the size-consistency issue we recall the main ideas of the method and how it is applied in practice.
Consider the Schr\"odinger equation in imaginary time for a system including $N$ particles with the {\it fixed-node} constraint, i.e. with the solution $\Phi({\bf R},t)$, where ${\bf R}$ is the electronic configuration and $t$ the time, forced to have the same nodal surface of some guiding function $\psi_G({\bf R})$ (the $3N-1$ hyper-surface where $\psi_G=0$). This is achieved, within the importance sampling scheme, by introducing the {\it mixed distribution} $f({\bf R},t) = \psi_G({\bf R})\Phi({\bf R},t)$, which satisfies the equation:
\begin{equation}\label{eqn:mixed}
- \frac{\partial f}{\partial t} =  -\frac{1}{2} \nabla^2 f +\nabla \cdot [{\bf V} f] - S  f \,.
\end{equation} 
Here we have omitted the functional dependence of the terms and 
${\bf V(R)} \equiv \nabla \log \left| \psi_G({\bf R}) \right|$ is 
usually called the {\em drift velocity}, 
%actually the local gradient of the guiding function; 
$S({\bf R}) \equiv E_T - E_L({\bf R})$ is the {\em branching} term, 
%$E_L({\bf R}) = H \psi_G({\bf R}) / \psi_G({\bf R})$ 
$E_L$ is the local energy, and $E_T$ is an energy shift. 
The three terms on the right hand side of Eq.~\ref{eqn:mixed} are responsible for diffusion, drift and branching processes, respectively.
Eq.~\ref{eqn:mixed} can be rewritten in integral form:
\begin{equation}\label{eqn:gf}
f({\bf R},t+t_0) = \int G({\bf R, R'};t) f({\bf R'},t_0) d{\bf R'}
\end{equation}
where $G({\bf R, R'};t)$
is the Green function for the importance sampling. 
The DMC method is a stochastic realization of Eq.~\ref{eqn:gf}, in which a series of {\it walkers} initially distributed as some $f({\bf R},0)=\sum_i \delta ({\bf r - r_i})$ is propagated ahead in time through a branching-drift-diffusion process, see Appendix~\ref{app:revDMC}. 
In the limit $t\to \infty$ the walkers end up distributed as $\psi_G({\bf R})\phi({\bf R})$, with $\phi({\bf R})$ the ground state of the Hamiltonian subject to the fixed-node constraint.

A practical implementation of Eq.~\ref{eqn:gf} faces a problem: 
$E_L({\bf R})$ and  ${\bf V}$ diverge at the nodes of $\psi_G$ as the inverse of the distance between the nodal surface and $\bf R$.
As $\tau\to0$ these two singularities are not an issue because the drift term prevents the walkers from approaching the nodal surface. However, for finite $\tau$, walkers can end up close to the nodal surface with catastrophic consequences. 
A practical solution to this problem is to introduce limits to the drift velocity and to the local energy.  
Umrigar, Nightingale and Runge~\cite{umrigar93} (UNR) proposed to replace 
${\bf V}=({\bf v}_1,\ldots,{\bf v}_N)$ with $\bar {\bf V}=(\bar{\bf v}_1,\ldots,\bar{\bf v}_N)$, defined as:
\begin{equation}\label{v}
{\bf \bar v}_i = \frac{-1+\sqrt{1+2av_i^2\tau}}{av_i^2\tau}{\bf v}_i; 
\quad
{\bf v}_i = \nabla_i \log \left| \psi_G({\bf R}) \right|,
\end{equation}
with $a$ an adjustable parameter between 0 and 1. This expression provides a rough approximation to the average velocity over a time-step, which has the effect of limiting the drift distance~\cite{umrigar93}.
The branching factor ${S}({\bf R})$ is replaced with:
\begin{equation}\label{eq:UNRbranch}
\bar{S}({\bf R}) = [E_T - E_{\rm best}] + [E_{\rm best} - E_L({\bf R})] \frac{{\bar V}}{V},  
\end{equation}
where $E_{\rm best}$ is the best estimate of the energy,  $V = \|{\bf V}\|$ and $\bar V = \| \bar {\bf V}\|$.
This limiting procedure is elegant and minimises instabilities because the divergences of $E_L({\bf R})$ at the nodes are cancelled by divergences in $V$. As a result it is now standard in most DMC simulations. 
However, this limiting procedure is an approximation of the Green function 
which renders DMC size-inconsistent, see discussion in Appendix~\ref{sec:sizecons}.

The issue disappears for $\tau \rightarrow 0$, where  $\bar{V}/V \rightarrow 1$, but for $\tau > 0$ the total energy is not proportional to the size of the system. 
Since the main application area of DMC is the calculation of medium to large systems for which relatively small energy differences are computed but very small $\tau$ cannot be afforded, this issue threatens the usefulness of DMC in material science. 

To quantify the size-consistency problem consider two systems $A$ and $B$ with energies $E_A$ and $E_B$, and define 
$E_{A,B}^{\rm separated}$ as the energy of the system with $A$ and $B$ at large enough distance from each other to have zero interaction.
The quantity $E_s = E_{A,B}^{\rm separated} - (E_{A} + E_{B})$ is therefore expected to be equal to zero and if it is not it measures the size-consistency error. To compute the binding energy of the system where $A$ and $B$ are interacting and have a total energy $E_{A,B}^{\rm bonded}$ it is useful to define $E_b = E_{A,B}^{\rm bonded} - (E_{A} + E_{B})$ and $E_{bs} = E_{A,B}^{\rm bonded} - E_{A,B}^{\rm separated}$. Here $E_b$ may be affected by a size-consistency problem, whereas $E_{bs}$ is not.
To illustrate the problem we have selected three representative examples 
with a broad range of interaction strengths, involving both isolated and periodic systems.

DMC simulations were carried out with the {\sc casino} code~\cite{casino}. We used Dirac-Fock pseudopotentials~\cite{trail05_NCHF, trail05_SRHF} with the locality approximation~\cite{mitas91}. The trial wavefunctions were of the Slater-Jastrow type with single Slater determinants and the single particle orbitals obtained from DFT-LDA plane-wave calculations performed with  {\sc pwscf}~\cite{pwscf}
and re-expanded in terms of B-splines~\cite{alfe04}. 

\begin{figure}
\includegraphics[width=3.3in]{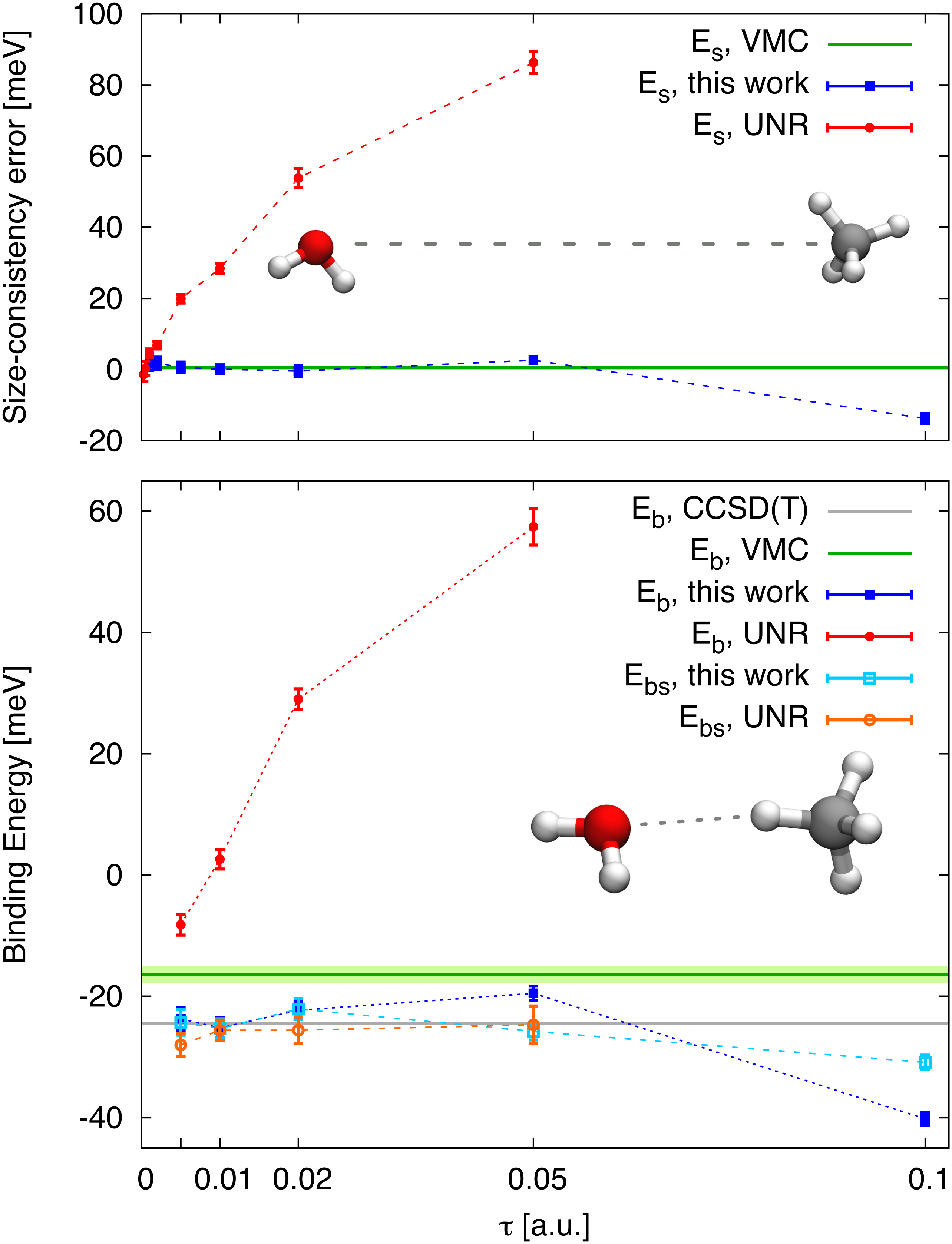}
\caption{\label{fig:FigMW} 
(Top) Size-consistency error $E_s$ (see text)
and 
(bottom)
binding energy 
[using two different definitions, $E_b$ and $E_{bs}$ (see text)]
for the CH$_4$-H$_2$O  system. 
% DMC
Results from the 
limited branching term given by Eq.~\ref{eq:UNRbranch} (UNR) 
or the approach introduced here Eqs.~\ref{eq:ZAbranch},\ref{eq:ecut} (this work) are reported.
VMC and CCSD(T)~\cite{gillan15} evaluations are also shown. Error bars are one standard deviation.
The insets show the structures of the complexes which have the molecules at large (top) and near the equilibrium (bottom) separation.
}
\end{figure}

Our first example is a system formed by a CH$_4$ ($A$) and a H$_2$O ($B$) molecule. 
$E_{A,B}^{\rm separated}$ is obtained for a C-O distance of 11.44~\AA.  
On the basis of CCSD(T) calculations we know that the residual interaction energy is $< 0.1$~meV, negligible for our purposes. 
$E_s$ is zero also for variational Monte Carlo (VMC), showing that the trial wavefunction of the dimer  $\psi_{\rm CH_4,H_2O}^{\rm separated}$ is effectively factorized: $\psi_{\rm CH_4,H_2O}^{\rm separated} = \psi_{\rm CH_4} \otimes\psi_{\rm H_2O}$.

In Fig.~\ref{fig:FigMW} (top) we plot $E_s$ computed with DMC as a function of $\tau$. 
For $\tau\to 0$, $E_s\to0$ as expected. 
However, at a typical time-step $\tau=0.005$~a.u.~\cite{gillan15} the error is already $\sim$20~meV, which is about the same size as the binding energy of the dimer near the equilibrium distance, and it grows to over 80~meV at $\tau=0.05$~a.u.. 
In  Fig.~\ref{fig:FigMW} (bottom) we show the 
binding energy of the molecule for a configuration near the equilibrium distance %CH$_4$-H$_2$O(bonded), 
\footnote{Note that this is not the water-methane dimer equilibrium configuration, but just a configuration in which the C-O distance is near the equilibrium value. }. 
As expected from the large size-consistency problem highlighted above, the binding energy computed with $E_b$ is wrong, and has a strong time-step dependence. Extrapolating to zero time-step using the whole $0.005 \le \tau \le 0.05$ range yields $E_b = 11\pm7$~meV.
Using only the range $0.005 \le \tau \le 0.02$ a value of $E_b = 21\pm2$~meV is obtained, which is close to the benchmark energy $E_b=24.5$~meV, obtained with 
coupled cluster with singles, doubles and perturbative triples 
(CCSD(T) and a large basis set)~\cite{gillan15}.  
By contrast, $E_{bs}$ is effectively time-step independent up to $\tau=0.05$, 
is in better agreement with the reference value, and removes the need for uncertain and arbitrary extrapolations.
The UNR limiting procedure is too unstable above $\tau=0.05$ and even at $\tau = 0.05$ we have not been able to obtain a very small statistical error due to instabilities in the simulation, see Appendix~\ref{app:instab}. 

Although one could envisage always using definitions analogous to $E_{bs}$ to compute binding energies, it is much more desirable to be able to use $E_b$ instead, particularly when one is concerned with the binding energy of more than just a dimer~\footnote{ For example, in the case of a cluster formed by a large number of molecules the construction of the system with all molecules far enough away from each other could be difficult, or even impossible, and alternative correction schemes would be required~\cite{gillan15}.  }.

To address this size-consistency issue we propose a new limiting procedure.
As proven in Appendix~\ref{sec:sizecons}, the UNR limit for the drift term, Eq.~\ref{v}, does not affect size-consistency, thus we only need to modify the branching term.
Our method is based on the idea that any modifications to the Green function should be as insensitive as possible  to the size of the system. 
%
%For the drift term, we keep using the limit of Eq.~\ref{v}, which improves the stability of the simulations and do not affect size-consistency~\footnote{ Umrigar recommends to set $a \le 0.5$~\cite{umrigar}. We find that our results depend weakly on the exact value of $a$. }.
Inspired by the prescriptions of DePasquale {\em et al.}~\cite{depasquale88}, in which the local energy entering  the branching factor is limited by a cutoff $E_{\rm cut}$, a modified branching factor is defined as:
\begin{eqnarray}\label{eq:ZAbranch}
{\bar S}({\bf R}) = E_T - {\bar E_L}({\bf R}); \nonumber  \\
{\bar E_L}({\bf R})  = E_{\rm best} + {\rm sign} [E_L{(\bf R)} - E_{\rm best}] \times \nonumber  \\
 \min\{E_{\rm cut}, |E_L{(\bf R)} - E_{\rm best}|\};  
\end{eqnarray}
In the original~\cite{depasquale88} recipe $E_{\rm cut} = 2/\sqrt{\tau}$. 
This has the consequence that for larger systems a larger fraction of the distribution of the branching factor is modified, leading again to a size-consistency issue.
Here we propose:
\begin{equation}\label{eq:ecut}
E_{\rm cut} = \alpha \sqrt{N/\tau},
\end{equation} 
where $N$ is the number of electrons in the system. Since the variance of the system is proportional to $N$, this  ensures that the proportion of the distribution of the branching factor modified by the cutoff is similar for systems with different values of $N$~\footnote{Note that, given $f(S_A)$ the distribution of the branching factor $S_A$ of some system $A$, the distribution $f(M;S_A)$ of a system containing $M$ non-interacting copies of $A$ does not have, in general, the same form. This is because the central limit theorem implies that $f(M;S_A)$ becomes Gaussian for large enough $M$, but in general $f(S_A)$ is not Gaussian. Thus the distribution cannot bemodified in a way that is exactly size-consistent and our proposed method is therefore only approximate.}. 
As with the original approach \cite{depasquale88}, the exact Green function is restored in the limit $\tau\to 0$.
The parameter $\alpha$ is an arbitrary constant to be conveniently chosen. For large enough values of $\alpha$ (and/or small values of $\tau$) the Green function becomes exact, but then singularities reappear. For small values of $\alpha$ (and/or large values of $\tau$) the bias in the DMC energy becomes large.  We have found that a good compromise is obtained by setting  $\alpha = 0.2$. 
The results obtained with this newly proposed scheme are displayed in Fig.~\ref{fig:FigMW}, showing that the bias in the DMC energy is now size-consistent up to very large values of $\tau$. The new scheme also reduces the time-step error on the absolute energies, see Appendix~\ref{app:MW}. 

If the composite system is made of non-identical subsystems (like our water-methane system) then the method becomes less accurate at large $\tau$, mainly because of the different widths of the $S$ distributions. In particular, the cutoff at $\tau = 0.1$~a.u. corresponds to $E_{\rm cut}$ of around $3.5~\sigma$, $2.7~\sigma$ and $3.0~\sigma$ for CH$_4$, H$_2$O and CH$_4$-H$_2$O, respectively, where $\sigma$ indicates the corresponding standard deviation of the VMC local energy~\footnote{
The standard deviation $\sigma_{\rm DMC}$ of the DMC distributions will, in general, be different from the $\sigma$ of the VMC distributions, but the same arguments would apply. }. 
With such small cutoff energies, the percentage of the respective distributions that are cut are different enough to affect the bias of the local energy in a non size-consistent way, which is why the error reappears at large values of $\tau$.

Binding energies computed with the new method are displayed in the bottom panel of Fig.~\ref{fig:FigMW}, showing that $E_{bs}$ has the same accuracy as that computed with the UNR branching factor, but now also $E_b$ is accurate. 
The new method is stable also for $\tau=$0.1~a.u., although at this very large value of the time-step the binding energy starts to show non negligible errors.
Note that in order to obtain a sufficiently high accuracy on $E_b$ with the UNR branching factor, without relying on extrapolations, we would need to reduce the time-step to at least $\tau \sim 0.0005$ a.u., which is two orders of magnitude smaller than what is required with our newly proposed method.

\begin{figure}
\includegraphics[width=3.3in]{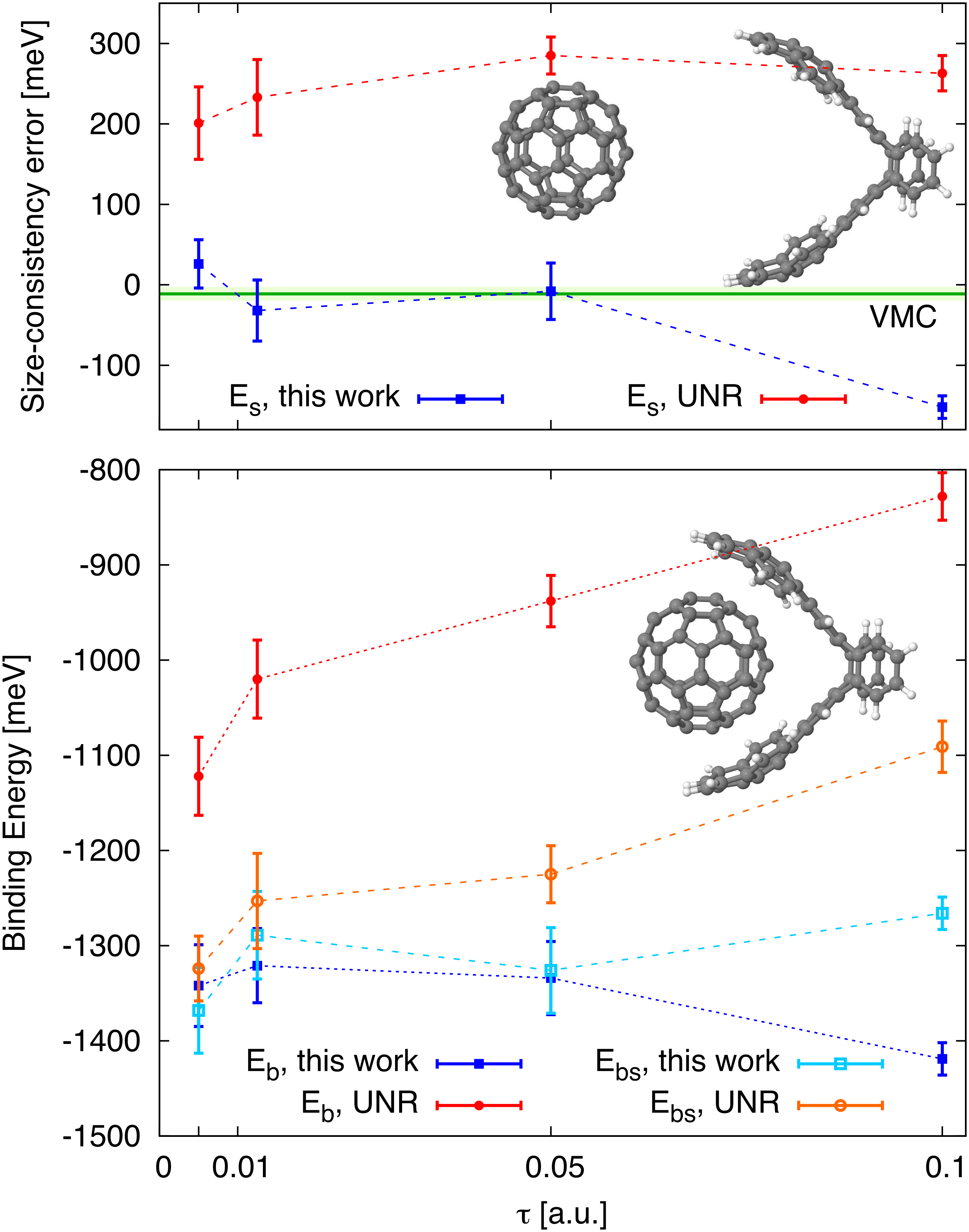}
\caption{\label{fig:Figbig}  
Same as Fig.~\ref{fig:FigMW} but in this case for the C$_{60}$-C$_{60}$H$_{28}$ system.}
\end{figure}

The second system we examined is the buckyball catcher, the C$_{60}$-C$_{60}$H$_{28}$ ($A-B$) complex. This is an example of a whole class of supramolecular systems which is 
generally out of reach of
the most accurate quantum chemistry methods and so at present DMC is the 
prime candidate for examining such systems.
For the calculation of $E_{A,B}^{\rm separated}$ we considered the system with the two fragments separated by 10~\AA. The residual interaction energy at this distance is $\simeq 10$~meV~\cite{jan}, which is again negligible compared to the energies involved. 
The new limiting procedure results in very good cancellation of time-step error and it is size-consistent up to at least $\tau=0.05$ a.u.. The UNR branching factor causes a slightly larger time-step dependence of both $E_b$ and $E_{bs}$, and the top panel of Fig.~\ref{fig:Figbig} highlights once again the size-consistency problem. Incidentally, the binding energy of this complex reported in~\cite{tkatchenko12} was  computed using UNR and $E_b$, therefore it had a size-consistency error of $\sim$ 0.2 eV.
Note that in this case any sensible extrapolation to zero time-step would result in a large size-consistency error, and therefore to obtain accurate results we should use $\tau\sim 0.0005$ a.u., if not even smaller, which is over two orders of magnitude more expensive and out of reach even on the biggest supercomputers currently available.%

\begin{figure}
\includegraphics[width=3.1in]{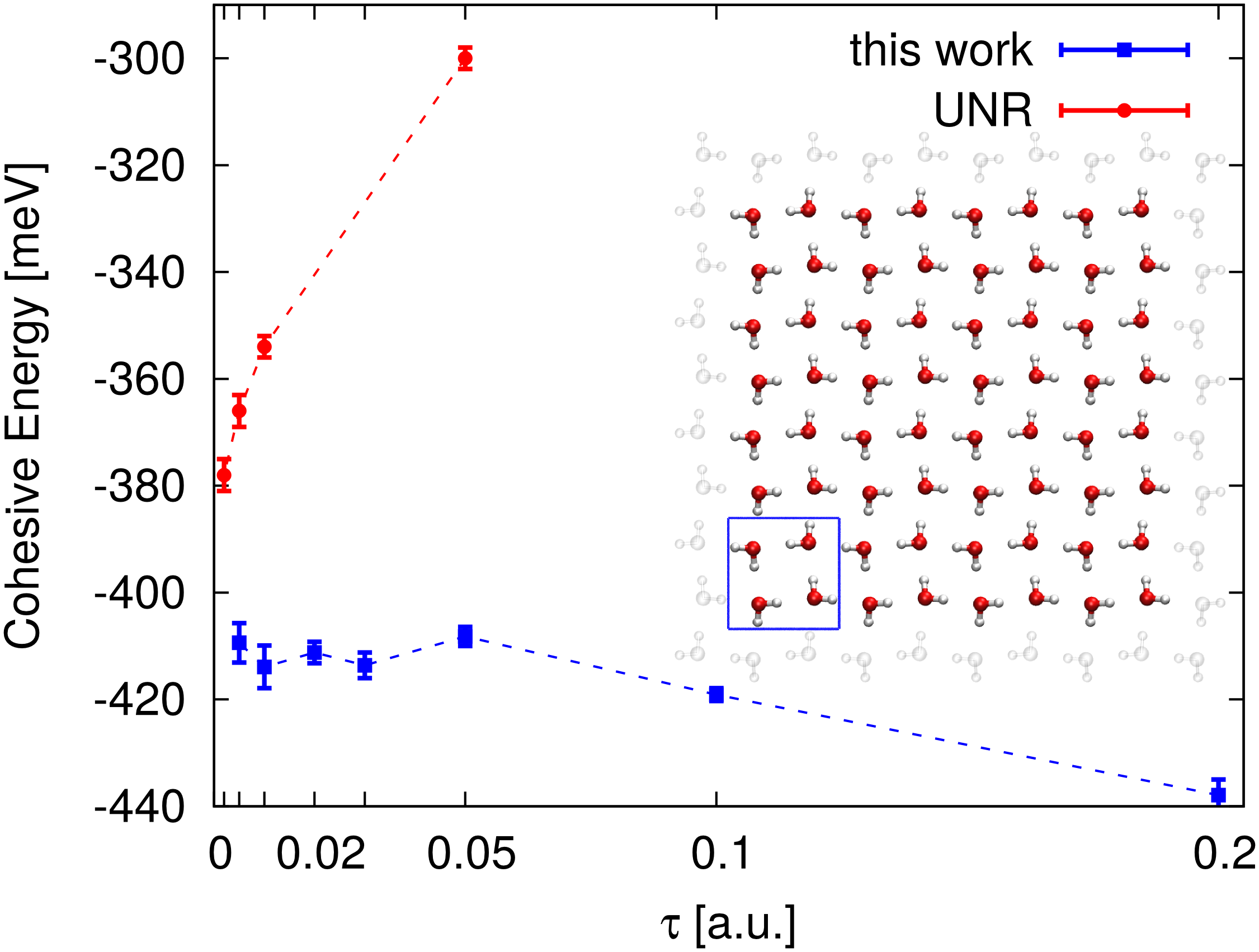}
\caption{\label{fig:ice} Cohesive energy of a two-dimensional periodic square ice system with the UNR and current branching terms. 
%comprised of 64 water molecules in the simulation cell. 
The inset of the structure shows the simulated 64 molecule supercell as colored molecules, and the primitive unit cell in the blue square. }
\end{figure}

Our third and final test was performed on a square lattice ice system, a H-bonded 2D-periodic system which has been the subject of recent theoretical~\cite{Chen:2016gz,Corsetti:2016da} and experimental~\cite{AlgaraSiller:2015cb} studies.
The simulation cell comprises 64 water molecules.
In Fig.~\ref{fig:ice} we show the cohesive energy as a function of time-step. The cohesive energy computed with the new limiting procedure is independent of time-step up to at least $\tau=0.05$ a.u., while that computed with the UNR branching factor has errors even at the shortest time-step that we could afford ($\tau=0.002$~a.u.). The non-linear trend of the UNR curve makes any $\tau\to 0$ extrapolation unreliable, unless simulations with $\tau<0.001$ a.u. could be afforded. 
Given the size of this system this makes such calculations prohibitively expensive.
Remarkably, the new method does not require any uncertain time-step extrapolations and yields a speedup of around two orders of magnitude.

\begin{figure*}
\includegraphics[width=7in]{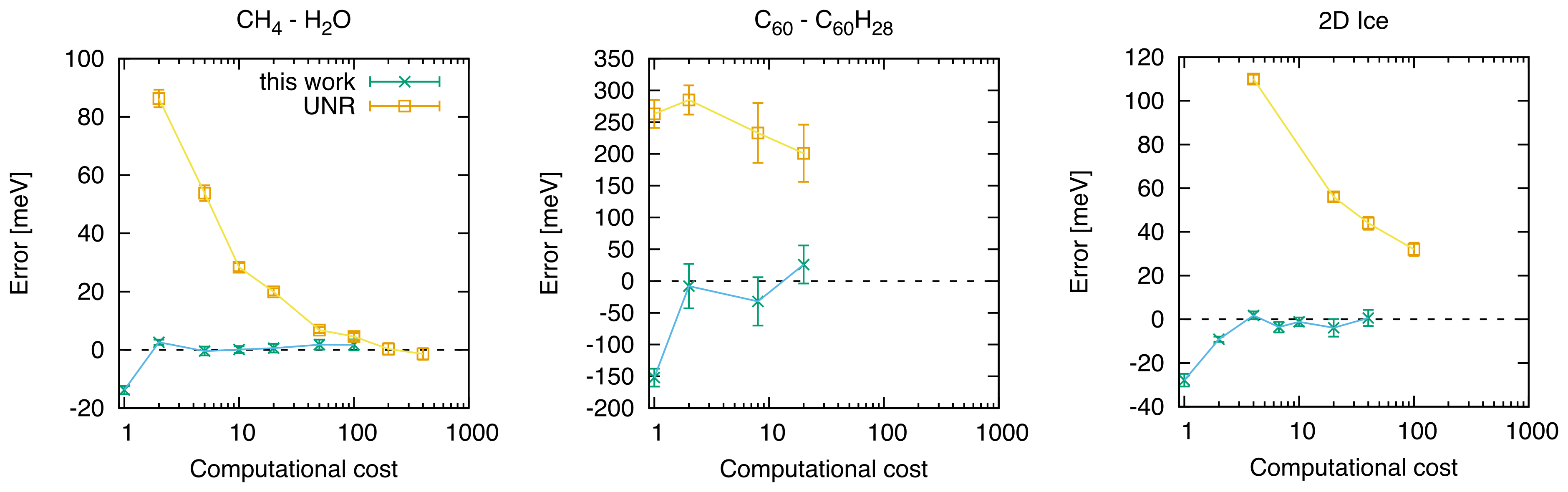}\\
\caption{\label{fig:accuracy} Size-consistency error  as a function of the computational cost in DMC calculations, for the three systems considered in the work, obtained by using UNR and present work prescriptions for the branching factor. Computational cost is in arbitrary units. }
\label{fig:speedup}
\end{figure*}

In summary, we have proposed a procedure that reduces DMC time-step errors by a large factor, and restores size-consistency. 
The method is based on the UMR scheme with an alternative branching factor. %with a modified  DePasquale {\em et al.} recipe. 
The modification is straightforward to implement,  requiring just a  change to a single line of code. We have demonstrated the new method on a CH$_4$-H$_2$O dimer, the C$_{60}$-C$_{60}$H$_{28}$ supramolecular system and 2-dimensional ice. 
Besides solving the size-consistency problem, speedups of two orders of magnitude are obtained (see Fig.~\ref{fig:speedup}) and the need for time-step extrapolations is removed.  
% new vistas:
The improvement appears particularly promising for investigations on molecular materials and to discriminate between crystal polymorphs.
Moreover, the recent emergence of QMC-based molecular dynamics~\cite{Mazzola:nat2014,Mazzola:prl2015,Zen:liqwat2015}, 
which until now has only been affordable within VMC, could now be in reach with the more accurate fixed-node DMC approach.

% acknowledgement
\begin{acknowledgements}
AZ and AM's work has been sponsored by the Air Force Office of Scientific Research, Air Force Material Command, USAF, under grant number FA8655-12-1-2099 and by the European Research Council under the European Union's Seventh Framework Programme (FP/2007-2013)/ERC Grant Agreement No. 616121 (HeteroIce project). AM is also supported by the Royal Society through a Wolfson Research merit Award. SS acknowledges  CINECA  for the use of  computational facilities, under IscrB\_SUMCHAL grant.
Calculations were performed on the U.K. national service ARCHER, the UK's national high-performance computing service, which is funded by the Office of Science and Technology through EPSRC's High End Computing Programme. This research also used resources of the Oak Ridge Leadership Computing Facility located in the Oak Ridge National Laboratory, which is supported by the Office of Science of the Department of Energy under Contract No. DE-AC05-00OR22725. We thank Cyrus Umrigar for useful discussions and Jan Hermann for providing the estimated residual binding energy of the  ${\rm C_{60}-C_{60}H_{28}(shifted)}$ complex.
\end{acknowledgements}

\appendix

\section*{Appendix}

In the first section of Appendix \ref{app:revDMC} we provide a short review of the DMC method, 
followed by a description of the DMC algorithm, the problem of the divergences in proximity of the nodal surface,
the instabilities in DMC simulations and the size-consistency issue met when  DMC  is stabilized by slightly modifying the algorithm. All this is used to contextualize the methodological improvements of this work.
Appendices \ref{app:MW}, \ref{app:big} and \ref{app:ice}  provide further details on the three examples shown in the paper.

\section{Review of DMC }\label{app:revDMC}

DMC energy evaluations are mostly concerned with the 
{\em mixed estimator}, defined as:
\begin{equation}\label{eq:Emix}
E_\textit{mix} = \frac{ \left< \phi \right| \hat H \left| \psi_G \right> }{ \left< \phi \right|  \left. \psi_G \right> }
\end{equation}
where 
$\psi_G$ is the {\em guiding function} 
(a parametrized wave function optimized within VMC schemes in order to be as close as possible to the ground state) and 
$\phi$ is the {\em exact ground state} wave function of the Hamiltonian $\hat H$.
As long as $\psi_G$ has a non-zero overlap with $\phi$, $E_\textit{mix}$ is equivalent to the {\em pure estimator} 
$\frac{ \left< \phi \right| \hat H \left| \phi \right> }{ \left< \phi \right|  \left. \phi \right>}$.

The exact wave function $\phi({\bf R})$ can be obtained from the solution 
$\Phi({\bf R},t)$ of the imaginary time Schr\" odinger equation
\begin{equation}\label{eqn:itse}
- \frac{\partial \Phi({\bf R},t)}{\partial t} =  -\frac{1}{2} \nabla^2 \Phi({\bf R},t) - (E_T - V_P({\bf R}))  \Phi({\bf R},t) 
\end{equation}
where $t$ is the time, ${\bf R}=({\bf r}_1, \ldots, {\bf r}_N)$ specifies the coordinates of the $N$ electrons,
$V_P$ is the potential energy and $E_T$ is an energy offset.
Given the boundary condition $\Phi({\bf R},0) = \psi_G({\bf R})$, for time $t \to \infty$ 
the imaginary time solution converges to the ground state:
$$
\lim_{t\to \infty} \Phi({\bf R},t) = \phi({\bf R}).
$$ 
It is often convenient to write the time evolution of $\Phi$ in terms of the Green function $G({\bf R \leftarrow R'};t)$:
\begin{equation}\label{eqn:gphi}
\Phi({\bf R},t_0+t) = \int G({\bf R \leftarrow R'};t) \Phi({\bf R'},t_0) d{\bf R'} .
\end{equation}
The Green function $G({\bf R \leftarrow R'};t)$, which satisfies an equation analogous to that of $\Phi$, prescribes how to propagate further in time the distribution $\Phi$.
Formally, we can write:
\begin{equation}\label{eqn:formalGbare}
G({\bf R \leftarrow R'};t) = \left< {\bf R} \right| e^{-t(\hat H - E_T)} \left| {\bf R'}  \right> \,.
\end{equation}
Unfortunately, $G({\bf R \leftarrow R'};t)$ is not exactly known for realistic systems.
However, by considering that the time interval $t$ can be divided in $n$ smaller intervals of time $\tau = t/n$,
and iteratively using Eq.~\ref{eqn:gphi} to write 
$\Phi({\bf R},t_{i})$ in terms of $\Phi({\bf R},t_{i-1})$, 
with 
$i=1,\ldots,n$ and
$t_i=t_0+i \tau$, 
we obtain the following expression for the Green function:
\begin{widetext}
\begin{equation}\label{eqn:expandGbare}
G({\bf R \leftarrow R'};t) = \int G({\bf R \leftarrow R}_1;\tau) \ldots G({\bf R}_{n-1} \leftarrow {\bf R'},\tau) d{\bf R}_1 \ldots d{\bf R}_{n-1} \,.
\end{equation}
\end{widetext}
For a small enough time step $\tau$,  the Green function can be approximated using the Trotter-Suzuki formula, which results in: 
\begin{equation}\label{eqn:appG1}
G({\bf R \leftarrow R'};\tau) \approx G_b({\bf R \leftarrow R'};\tau) G_d({\bf R \leftarrow R'};\tau)
\end{equation}
where 
$$
G_d({\bf R \leftarrow R'};\tau) =  (2 \pi \tau)^{-\frac{3}{2}  N } \exp\left[ - \frac{ ( {\bf R - R'} )^2 }{ 2 \tau } \right] 
$$
is a diffusion term, and 
$$
G_b({\bf R \leftarrow R'};\tau) = \exp\left[ \tau \frac{ 2 E_T - V_P({\bf R}) - V_P({\bf R'}) }{ 2 } \right] 
$$
is a branching term.
The DMC algorithm is a stochastic realization of Eq.~\ref{eqn:gphi}, in which a series of {\it walkers} initially distributed as some $\Phi({\bf R},0)$ %=\sum_\alpha \delta ({\bf R - R}^\alpha)$ 
is propagated ahead in time with the short time approximation to the Green function in Eq.~\ref{eqn:appG1}. In the long time limit the walkers become distributed as $\phi({\bf R})$. 

The method works perfectly well for bosons, as the ground state of the Hamiltonian is node-less. However, the fermionic ground state is generally difficult to calculate, because it is an excited state of the Hamiltonian. The difficulty comes from the fact that in the time evolution of Eq.~\ref{eqn:itse} the weight of the ground state becomes exponentially
dominant compared to excited states, and so the fermionic signal is quickly lost into noise.
The common solution is to embrace the {\em fixed node approximation}: $\Phi({\bf R},t)$ in constrained to have the same nodal surface of some guiding function $\psi_G({\bf R})$. The constraint makes DMC only approximate, and the variational principle then implies that the fixed-node DMC energy is an upper bound of the true fermionic ground state energy. If the nodal surface of the guiding function is exact then also the fixed-node DMC energy is exact.

The fixed-node constraint is conveniently implemented by introducing the mixed distribution $f({\bf R},t) = \psi_G({\bf R}) \Phi({\bf R},t)$,
which satisfies the equation:
\begin{equation}\label{eqn:mix}
- \frac{\partial f({\bf R},t)}{\partial t} =  -\frac{1}{2} \nabla^2 f({\bf R},t) +\nabla \cdot [{\bf V}({\bf R}) f({\bf R},t)] - S({\bf R})  f({\bf R},t)
\end{equation}
(see Eq.~\ref{eqn:mixed}), where 
${\bf V(R)} \equiv \nabla \log \left| \psi_G({\bf R}) \right|$ is the {\em drift velocity}, or local gradient, and
$S({\bf R}) \equiv E_T - E_L({\bf R})$ is the {\em branching} term, with $E_L({\bf R}) = \psi_G({\bf R})^{-1} \hat H \psi_G({\bf R})$ the {\em local energy}.
Note that in Eq.~\ref{eqn:mix} there is an additional drift term that was not present in the original imaginary time Schr\" odinger equation for $\Phi$. 
The mixed distribution $f$ has the border condition
$
f({\bf R},0) = \psi_G({\bf R})^2
$
and, in the limit of large time $t$:
$$
\lim_{t\to \infty} f({\bf R},t) = \psi_G({\bf R}) \phi({\bf R}).
$$
Thus, the mixed estimator can be written as:
\begin{equation}\label{eq:Emix2}
E_\textit{mix} = \lim_{t\to \infty} \frac{ \int E_L({\bf R}) f({\bf R},t) d{\bf R} }{ \int f({\bf R},t) d{\bf R} }.
\end{equation}

It is convenient to write the time evolution of $f$ in terms of the Green function $\tilde G({\bf R \leftarrow R'};t)$, which prescribes  how to propagate further in time the distribution $f$:
\begin{equation}\label{eqn:gf}
f({\bf R},t_0+t) = \int \tilde G({\bf R \leftarrow R'};t) f({\bf R'},t_0) d{\bf R'} ,
\end{equation}
where $\tilde G({\bf R \leftarrow R'};t)$ satisfies an equation analogous to that of $f$, and formally can be written as:
\begin{equation}\label{eqn:formalG}
\tilde G({\bf R \leftarrow R'};t) = \frac{\psi_G({\bf R})}{\psi_G({\bf R'})} \left< {\bf R} \right| e^{-t(\hat H - E_T)} \left| {\bf R'}  \right> \,.
\end{equation}
Again, $\tilde G({\bf R \leftarrow R'};t)$ is not exactly known for realistic systems, but we can use the same trick of splitting $t$ in $n$ time steps of length $\tau=t/n$.
We obtain the following expression for the Green function:
\begin{widetext}
\begin{equation}\label{eqn:expandG}
\tilde G({\bf R \leftarrow R'};t) = \int \tilde G({\bf R \leftarrow R}_{n-1};\tau) \ldots \tilde G({\bf R}_{1} \leftarrow {\bf R'},\tau) d{\bf R}_1 \ldots d{\bf R}_{n-1} \,.
\end{equation}
For a small enough time step $\tau$, 
$\tilde G({\bf R}_i, {\bf R}_{i+1};\tau)$
is approximated by the Green functions for purely drift, diffusion and branching processes.
This leads to:
\begin{equation}\label{eqn:appG}
\tilde G({\bf R \leftarrow R'};\tau) \approx \tilde G_b({\bf R \leftarrow R'};\tau) \tilde G_d({\bf R \leftarrow R'};\tau)
\end{equation}
where 
$$
\tilde G_d({\bf R \leftarrow R'};\tau) =  (2 \pi \tau)^{-\frac{3}{2}  N } \exp\left[ - \frac{ ( {\bf R - R' - \tau V(R')} )^2 }{ 2 \tau } \right] 
$$
is the drift-diffusion term, 
and 
$$
\tilde G_b({\bf R \leftarrow R'};\tau) = \exp\left[ \tau \frac{ S({\bf R}) + S({\bf R'}) }{ 2 } \right] 
$$
is the branching term.
\end{widetext}

Eq.~\ref{eqn:mix} also introduces {\em importance sampling}.
Beside concentrating the sampling in the important part of the phase space, an additional
advantage of importance sampling over simple sampling is that the branching term depends on the local energy $E_L({\bf R})$, and not on the potential energy $V_P({\bf R})$. Since $E_L({\bf R})$ is much smother than $V_P({\bf R})$, and it is constant in the limit of $\psi_G \sim \phi$,  the stability of the DMC simulation is greatly  enhanced. 
The error on this approximate expression for $\tilde G({\bf R}_i, {\bf R}_{i+1};\tau)$ can be evaluated using the
Zassenhaus formula~\cite{Suzuki:1977bw},
and the leading correction is of order ${\cal O}(\tau^2)$.
This translates into an error of order ${\cal O}(\tau)$ on $\tilde G({\bf R, R'};t)$ (see Eq.~\ref{eqn:expandG}).
In the limit of $ \tau \rightarrow 0$ the error on the Green function is zero, but the computational cost is $\propto 1/\tau$ because $\tilde G_b({\bf R \leftarrow R'};t)$ is split in $n=t/\tau$ terms.

\subsection{ DMC algorithm }
% how DMC in practice
We discuss here  how the DMC algorithm actually works.
At each time $t$ the distribution $f({\bf R},t)$ can be represented by a discrete set $\{ {\bf R}^\alpha(t),w^\alpha(t) \}_{\alpha=1,\ldots,n_w(t)}$ of walkers ({\em i.e.} sampling points ${\bf R}^\alpha$ with a weight $w^\alpha$),
such that 
$f({\bf R},t) \sim { \sum_\alpha w^\alpha \delta({\bf R} - {\bf R}^\alpha) / \sum_\alpha w^\alpha }$.
By using the Metropolis algorithm we can easily generate an ensemble of configurations 
$\{ {\bf R}^\alpha \}_{\alpha=1,\ldots,n_w}$ ({\em i.e.}, a set of walkers with unit weight) that correspond to the initial distribution $f({\bf R},0)= \psi_G({\bf R}_n)^2$.
In DMC we need to project forward in time the walkers in order to calculated the mixed distribution for $f({\bf R},t\to \infty)$.

If in Eq.~\ref{eq:Emix2} we 
express the mixed distribution $f({\bf R},t)$ as in Eq.~\ref{eqn:gf} (with initial distribution $f({\bf R},0) = \psi_G({\bf R})^2$),
and we  expand the Green function as in Eq.~\ref{eqn:expandG} (with $t=n\tau$),
we obtain that the mixed estimator is rewritten in the following way: 
\begin{widetext}
\begin{equation}\label{eqn:Emix3}
E_\textit{mix} = \lim_{n\to \infty} \frac{ 
	\int E_L({\bf R}_n)  
		\tilde G({\bf R}_n \leftarrow {\bf R}_{n-1};\tau) \ldots 
		\tilde G({\bf R}_{1} \leftarrow {\bf R}_0,\tau) \psi_G({\bf R}_n)^2  
		d{\bf R}_0 \ldots d{\bf R}_n 
	}{ 
	\int 	
		\tilde G({\bf R}_n \leftarrow {\bf R}_{n-1};\tau) \ldots 
		\tilde G({\bf R}_{1} \leftarrow {\bf R}_0,\tau) \psi_G({\bf R}_n)^2  
		d{\bf R}_0 \ldots d{\bf R}_n 
	} \, ,
\end{equation}
and using the approximation in Eq.~\ref{eqn:appG} for the Green function with small $\tau$ we have:
\begin{equation}\label{eqn:Emix4}
E_\textit{mix} \simeq \lim_{n\to \infty} \frac{ 
	(2 \pi \tau)^{-\frac{3}{2} n N }
	\int 
	E_L({\bf R}_n)  
		\prod_{i=0}^{n-1} \left\{
		\exp\left[ - \frac{ ( {\bf R}_{i+1} - {\bf R}_i - \tau {\bf V}({\bf R}_i) )^2 }{ 2 \tau } \right] 
		 \exp\left[ \tau \frac{ S({\bf R}_{i+1}) + S({\bf R}_i) }{ 2 } \right] 
		 \right\}
		\psi_G({\bf R}_n)^2  
		d{\bf R}_1 \ldots d{\bf R}_n 
	}{ 
	(2 \pi \tau)^{-\frac{3}{2} n N }
	\int 
		\prod_{i=0}^{n-1} \left\{
		\exp\left[ - \frac{ ( {\bf R}_{i+1} - {\bf R}_i - \tau {\bf V}({\bf R}_i) )^2 }{ 2 \tau } \right] 
		 \exp\left[ \tau \frac{ S({\bf R}_{i+1}) + S({\bf R}_i) }{ 2 } \right] 
		 \right\}
		\psi_G({\bf R}_n)^2  
		d{\bf R}_1 \ldots d{\bf R}_n 
	} \, .
\end{equation}
\end{widetext}
Thus, according to the RHS of Eq.~\ref{eqn:Emix4}, each walker 
%$\{{\bf R}^\alpha_i,w^\alpha_i\}$ at time $t=i*\tau$ 
evolves  in time according to a branching-drift-diffusion process: 
given the configuration ${\bf R}^\alpha_i$ and weight $w^\alpha_i$ at time $t=i*\tau$,  the walker drift-diffuse as follows: 
\begin{equation}\label{eq:driftdiff}
{\bf R}^\alpha_{i} \to
{\bf R}^\alpha_{i+1} = {\bf R}^\alpha_i + \tau {\bf V}({\bf R}^\alpha_i) + \sqrt{\tau} { \eta} \,,
\end{equation}
where ${ \eta}$ is a $3N$-dimensional random vector generated from a normal distribution with zero mean and unit variance, and the walker weight evolves as:
\begin{equation}\label{eq:weight}
w^\alpha_{i} \to
w^\alpha_{i+1} = w^\alpha_i * \exp\left[ \tau \frac{ S({\bf R}^\alpha_{i+1}) + S({\bf R}^\alpha_i) }{ 2 } \right]  \,.
\end{equation}
The evolution of the weight is efficiently realized by using a branching (birth/death) algorithm, where walkers with small weight are killed and walkers with high weight are replicated \cite{foulkes01}.
Moreover, a Metropolis acceptance/rejection move is usually introduced after the drift-diffusion step\cite{Reynolds:1982,umrigar93}, in order to satisfy the detailed balance and reduce the time-step error, and with that an efficient time-step $\tau_\textrm{\small eff}$, which rescales the nominal time-step $\tau$ taking into account  the acceptance probability, is used in Eq.~\ref{eq:weight} in place of $\tau$.

Finally, given the chosen time-step $\tau$ and a sufficiently large number $n$ of DMC steps, the mixed energy is calculated as:
\begin{equation}\label{eqn:Emix5}
E_\textit{mix}^\tau = 
\frac{ 
\left< E_L({\bf R}_{n}^\alpha) w_{n}^\alpha \right>_\alpha
}{ \left< w_{n}^\alpha \right>_\alpha } \,, 
\end{equation}
where $\left< \cdot \right>_\alpha$ is the average over all the walkers.
% actual DMC
Clearly, this evaluation is affected by a stochastic error inversely proportional to the square root of the number $n_w$ of walkers. 
In order to increase the precision of the evaluations it is not necessary to use a huge number of walkers; it is much more efficient, because of the equilibration time, to propagate further in time the walkers and to use the following expression to evaluate the mixed energy:
\begin{equation}\label{eqn:actualEmix}
E_\textit{mix}^\tau = 
{1 \over M} \sum_{m=1}^M
\frac{ 
\left< E_L({\bf R}_{n+m}^\alpha) w_{n+m}^\alpha \right>_\alpha
}{ \left< w_{n+m}^\alpha \right>_\alpha } \,.
\end{equation}
Notice that in Eq.~\ref{eqn:actualEmix} the walkers provide almost independent evaluations, but the local energies are instead serially correlated, with a correlation time proportional to $\tau$. Thus, in evaluating the stochastic error for the mixed energy it is important to get rid of the serial correlation, for instance by using the ``blocking method'' \cite{flyvbjerg_error_1989}.
Sometimes the  estimator actually used can be slightly different from Eq.~\ref{eqn:actualEmix} -- for instance some corrections are sometimes introduced in order to correct for the finite population bias ({\em i.e.}, having a finite number of walkers can introduce a  bias) -- but the size-consistency issue here addressed is unaffected by these corrections.

\subsection{Divergences in proximity of the nodal surface}

Close to the nodal surface $\Sigma_G$ of the guiding function $\psi_G$ the approximation in Eq.~\ref{eqn:appG} is problematic, because a configuration ${\bf R}$ at a distance $\delta$ from $\Sigma_G$ has both the local gradient $\bf V(R)$ and the local energy $E_L({\bf R})$ (and consequently the branching term $S({\bf R})$) diverging in modulus as $1/\delta$, leading to instabilities and big finite time step errors.
This problem has been tackled both by \citet{depasquale88} and \citet{umrigar93}, who proposed modifications for
$\bf V(R)$ and for $S({\bf R})$ for $\bf R$ close to $\Sigma_G$ to eliminate these divergences. 
These modifications are strictly related to  the size-inconsistency issue addressed in this work.

\subsection{DMC instabilities}\label{app:instab}

DMC instabilities are  uncontrolled walker population fluctuations ({\em i.e.}, weights $w^\alpha_i$ experiencing huge changes in a single step $i\to i+1$, see Eq.~\ref{eq:weight}), which jeopardize the DMC energy evaluations and makes the simulation unfeasible. They are mainly due to walkers reaching regions of diverging local energy (because of the pseudo-potential or proximity to the nodal surface), and in particular for $E_L({\bf R}) \to -\infty$ the branching term $S({\bf R})$ leads to proliferation of walkers from just one problematic configuration. 
Instabilities are strictly related with time step $\tau$: with small $\tau$ instabilities are usually under control, but as larger and larger values of $\tau$ are considered instabilities are more often observed. 
The reason is that the diffusion step is random and proportional to $\sqrt{\tau}$, see Eq.~\ref{eq:driftdiff}, and if the time-step is too large there is some chance to fall into the problematic regions, because the drift step %, proportional to $\tau$,
is unable to keep electrons away for the divergences.
A small enough $\tau$ allows the drift step to recover from a ``bad'' diffusion step.
% UNR
As a matter of fact, DMC simulations with no modifications to the drift and branching terms are stable only for tiny values of $\tau$, making schemes as those proposed by \citet{depasquale88} or \citet{umrigar93} necessary in actual calculations, the latter being much more stable than the former. 
The new limiting scheme proposed  in this work 
(which is the same of \citet{umrigar93} for the drift, Eq.~3 of the letter, and the one in Eq.~5 of the letter for the branching) appears as effective as the limiting scheme of  \citet{umrigar93} (see Eqs.~3 and 4 of the letter), if not better, in keeping the DMC simulation stable.

A pragmatic way to recover from a diverging population count (population explosion) is to back-track the simulation to a region far from the instability, run the random number generator idle for a number of cycles, and resume the DMC simulation. Often this procedure sends the simulation to a different region of phase space, avoiding the instability. However, if the instabilities are too frequent, the simulation becomes impractical or even impossible. To highlight the improvement in the stability of the calculations using the new limiting procedure, consider for example the CH$_4$ - H$_2$O dimer in the bound configuration. Using the UNR limiting procedure and $\tau= 0.05$ a.u. we encountered 32 population explosions in $\sim 26,000$ steps (population size: 20,480 walkers). No simulations were possible with any larger value of time step. By contrast, using the new limiting procedure we observed no instabilities in $\sim 176,000$ steps at $\tau= 0.05$ a.u., and also no instabilities in $\sim 250,000$ steps at $\tau= 0.1$ a.u..

\subsection{ Size-consistency in DMC }\label{sec:sizecons}

As discussed in the letter, a method is size-consistent if the energy $E_{AB}$ of any system $AB$ constituted by the two non-interacting subsystems $A$ and $B$, is equal to the sum $E_A+E_B$ of the energies of  individual  subsystems.
%As in the letter, we assume here to deal with closed shell subsystems, so that they can be described with a single Slater determinant (and also with a Jastrow correlated single Slater determinant).  
As in the letter, we assume here to deal with systems that are size-consistent when described with a single Slater determinant (so, also with a Jastrow correlated single Slater determinant).
In this section we show that the fixed-node DMC with importance sampling ({\em i.e.}, Eq.~\ref{eqn:Emix5}) is size-consistent for any $\tau$, but if the modifications to the branching proposed by \citet{umrigar93} are used DMC is size-consistent only in the limit of $\tau\to 0$.

% premesse
Clearly, any configuration ${\bf R}^{[AB]}$ of the systems $AB$ is given by the configurations ${\bf R}^{[A]} $ and ${\bf R}^{[B]} $ of the subsystems $A$ and $B$, because any electron in AB belongs either to the subsystem $A$ or to $B$.
Mathematically, this means that the vectorial space where the configurations ${\bf R}^{[AB]}$ live is the direct sum of the two vectorial spaces where  ${\bf R}^{[A]} $ and ${\bf R}^{[B]} $ live, and we can write (with a little abuse of notation): 
\begin{equation}\label{eq:Rab}
{\bf R}^{[AB]} = {\bf R}^{[A]} \oplus {\bf R}^{[B]} \,.
\end{equation}
As discussed in the letter, the guiding wave function factorizes, {\em i.e.}: 
\begin{equation}\label{eq:psiab}
\psi_G^{[AB]}({\bf R}^{[AB]}) = \psi_G^{[A]}({\bf R}^{[A]}) \otimes \psi_G^{[B]}({\bf R}^{[B]}) 
\end{equation}
whenever $A$ and $B$ are far away. 
From the properties of the hamiltonian operator it follows that the local energy is additive:
\begin{equation}\label{eq:elocadd}
E_L^{[AB]}({\bf R}^{[AB]}) = E_L^{[A]}({\bf R}^{[A]}) + E_L^{[B]}({\bf R}^{[B]}) \,,
\end{equation}
which proves that VMC is size-consistent.
Moreover, considering that the drift velocity is the local gradient, it is easy to show that:
\begin{equation}\label{eq:driftadd}
{\bf V}^{[AB]}({\bf R}^{[AB]}) = {\bf V}^{[A]}({\bf R}^{[A]}) \oplus {\bf V}^{[B]}({\bf R}^{[B]}) \,,
\end{equation}
where the symbol $\oplus$ is used in the same way as in Eq.~\ref{eq:Rab}.

In order to address the properties of the DMC mixed energy $E_\textit{mix}^\tau$ evaluated for a finite value $\tau$ of the time-step, we can consider Eq.~\ref{eqn:Emix5}.
According to Eq.~\ref{eq:weight}, the weight is  
$w_{n}^\alpha = \exp{\left[ \tau \sum_{i}^n S({\bf R}_{i}^\alpha) \right]}$ (here, for simplicity, we have slightly simplified the expression, neglecting that the first and last step have a weight that is 1/2)
and including that the branching term $S({\bf R}_{i}^\alpha) = E_T - E_L({\bf R}_{i}^\alpha)$, it is straightforward to see that:
\begin{equation}
\label{eqn:Emix6}
E_\textit{mix}^\tau = 
\frac{ 
\left< E_L({\bf R}_{n}^\alpha) e^{ - \tau \sum_{i}^n E_L({\bf R}_{i}^\alpha) } \right>_\alpha
}{ \left< e^{ - \tau \sum_{i}^n E_L({\bf R}_{i}^\alpha) }  \right>_\alpha } \,.
\end{equation}
By using Eq.~\ref{eqn:Emix6}, the additivity of the local energy (Eq.~\ref{eq:elocadd}) and of the drift velocity (Eq.~\ref{eq:driftadd}), and some algebra, it is easy to prove that:
\begin{widetext}
\begin{equation}\label{eqn:SCdmc}
\textrm{DMC with no modifications:} \qquad
{E_\textit{mix}^\tau}^{[AB]} = 
{E_\textit{mix}^\tau}^{[A]} + {E_\textit{mix}^\tau}^{[B]} 
\,,
\end{equation}
for any value of the time-step $\tau$, and of course also for $\tau\to 0$.
The main point of the proof is that the additivity of the local energy imply the factorization of the weight, {\em i.e.}:
\begin{equation}
e^{ - \tau \sum_{i}^n E_L^{[AB]}({{\bf R}_{i}^\alpha}^{[AB]})} = 
e^{ - \tau \sum_{i}^n E_L^{[A]}({{\bf R}_{i}^\alpha}^{[A]})} * 
e^{ - \tau \sum_{i}^n E_L^{[B]}({{\bf R}_{i}^\alpha}^{[B]})} 
\qquad \textrm{for any walker } \alpha.
\end{equation}
In principle, it could be explicitly tested that DMC with no modifications satisfy the size-consistency for any finite time-step, but in practice it can be done only for very small values of $\tau$ because of the instabilities discussed in Section~\ref{app:instab}. 

% UNR 
The UNR modification to the drift, as reported in Eq.~3 of the letter, does not affect the additivity of the drift (because the correction is performed independently for each electron), and we have that:
\begin{equation}\label{eq:UNRdriftadd}
\textrm{DMC with UNR modifications:} \qquad
{\bf \bar V}^{[AB]}({\bf R}^{[AB]}) = {\bf \bar V}^{[A]}({\bf R}^{[A]}) \oplus {\bf \bar V}^{[B]}({\bf R}^{[B]}) \,.
\end{equation}
which clearly does not affect the size-consistency of the method.
The source of the size-inconsistency is instead the UNR modification to the branching term, see Eq.~4 in the letter, because we have that: 
\begin{equation}\label{eq:UNRbranchadd}
\textrm{DMC with UNR modifications:} \qquad
{\bar S}^{[AB]}({\bf R}^{[AB]}) \neq {\bar S}^{[A]}({\bf R}^{[A]}) + {\bar S}^{[B]}({\bf R}^{[B]}) \,.
\end{equation}
because of the term ${\bar V / V}$ appearing in the expression of $\bar S$.
This imply that the weight of a DMC realization does not factorize any more, that is:
\begin{equation}\label{eqn:UNRweight}
\textrm{DMC with UNR modifications:} \qquad
e^{ \tau \sum_{i}^n \bar S^{[AB]}({{\bf R}_{i}^\alpha}^{[AB]})} \neq 
e^{ \tau \sum_{i}^n \bar S^{[A]}({{\bf R}_{i}^\alpha}^{[A]})} * 
e^{ \tau \sum_{i}^n \bar S^{[B]}({{\bf R}_{i}^\alpha}^{[B]})} 
\,.
\end{equation}
However, in the limit of $\tau \to 0$ we have that $\bar V \to V$ and $\bar S \to S$, thus UNR approaches asymptotically the case of no modifications, where size-consistency is proven.

% this work
The scheme proposed in this work (named here ZSGMA, from authors' names), see Eqs.~5 and 6 in the letter, is exactly size-consistent for $E_\textrm{cut}\to \infty$ (namely, for $\alpha\to\infty$ or $\tau\to0$), because the branching $\bar S$ becomes equivalent to $S$, which factorizes exactly, so we recover the unmodified DMC algorithm.
The method is only approximated for finite $E_\textrm{cut}$;
the modified branching term is not exactly additive, {\em i.e.}
${\bar S}^{[AB]}({\bf R}^{[AB]}) \neq \sum_{i}^n {\bar S}^{[A]}({\bf R}^{[A]}) + \sum_{i}^n {\bar S}^{[B]}({\bf R}^{[B]})$,
but what we approximatively satisfy is that:
\begin{equation}\label{eq:ZSGMAbranch}
\textrm{DMC with ZSGMA modifications:} \qquad
\sum_{i}^n {\bar S}^{[AB]}({\bf R}^{[AB]}) \sim
\sum_{i}^n {\bar S}^{[A]}({\bf R}^{[A]}) + \sum_{i}^n {\bar S}^{[B]}({\bf R}^{[B]}) \,,
\end{equation}
at least when $E_\textrm{cut}$ is large enough.
This happens because, assuming that $E_T$ is properly set,
we have that $\bar S$ can be seen as a  random variable of zero mean and a variance proportional to $\sqrt{N}$.
In order to satisfy Eq.~\ref{eq:ZSGMAbranch}, at least approximatively, we require that the number of times we perform a cut on $\bar S$ is independent on the size of the system and with a random sign. This implyes a value of $E_\textrm{cut} \propto \sqrt{\textrm{VAR}(\bar S)}$.
\end{widetext}

\newpage

\section{Water-Methane dimer}\label{app:MW}

\begin{figure}[ht]
\includegraphics[width=3.5in]{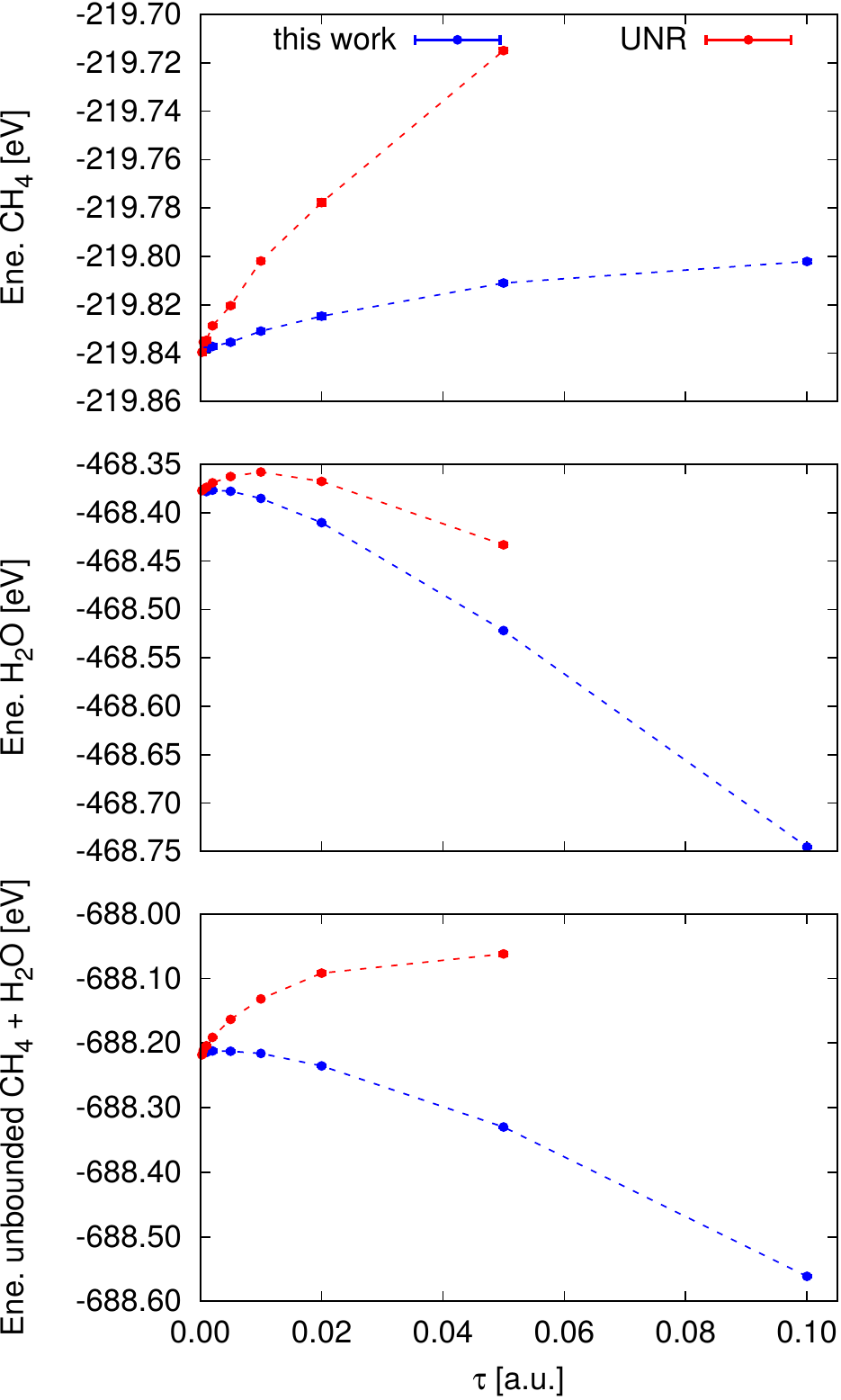}
\caption{\label{fig:Fig2} Energies of the 
CH$_4$ (top panel), 
H$_2$O (middle), 
and unbounded CH$_4$-H$_2$O (bottom) 
systems as function of time step $\tau$, calculated using the UNR and present work prescriptions for the limits on the branching factor. Error bars are smaller than the size of the symbols. }
\end{figure}

In Fig.~\ref{fig:Fig2} we display the energy of the dimer, $E_{\rm CH_4-H_2O(\rm shifted)}$ as well and the energies of the monomers, $E_{\rm CH_4}$ and $E_{\rm H_2O}$, computed in independent calculations performed with simulation cells containing either the CH$_4$-H$_2$O(\rm shifted) dimer or the isolated CH$_4$ and H$_2$O monomers, respectively. 

Single particle wavefunctions were obtained using a plane-wave cutoff of 300 Ry, and re-expanded in terms of B-splines with the natural grid spacing $a=\pi/G{\rm max}$, where $G_{\rm max}$ is the magnitude of the largest plane wave in the expansion.  
The Jastrow factor used in the trial wavefunction of the system included a two-body electron-electron (e-e) term; three different two-body electron-nucleus (e-n) terms for C, O and H, respectively; and three different three-body electron-electron-nucleus (e-e-n) terms, for C, O and H. Of course, for the isolated CH$_4$ and H$_2$O systems we only included the e-n and the e-e-n terms for C, H and O, H, respectively, but a part form this difference the Jastrow factors were exactly the same in all systems. The cutoff radii of the e-e, e-n, and e-e-n terms were all lower than 3.5~\AA, and the large distance between the two molecules  guarantees that the overlap between their respective orbitals is effectively zero. Therefore the trial wavefunction of the dimer  $\psi_{\rm CH_4-H_2O(shifted)}$, is effectively the appropriately antisymmetrised product of the trial wavefunctions $\psi_{\rm CH_4}$ and $\psi_{\rm H_2O}$  of the CH$_4$ and the H$_2$O sub-systems, respectively: $\psi_{\rm CH_4-H_2O(shifted)} = \psi_{\rm CH_4} \otimes\psi_{\rm H_2O}$. The variances of the local energy with the variational Monte Carlo (VMC) distributions were  $\sim 0.72, 0.26$ and $0.45$ Ha$^2$ for the CH$_4$-H$_2$O, CH$_4$ and H$_2$O systems, respectively.

As seen in the paper, the finite time-step error in the binding energy, whenever the $E_b$ evaluation is used, is mostly due to the size consistency error.
The speedup obtained by using  present work prescriptions for the branching factor in comparison with UNR branching factor is of two orders of magnitude, as it is shown in Fig.~\ref{fig:accuracy}(left). In this system there is the possibility to use $E_{bs}$ and to alleviate the size-consistency issue of the UNR prescription for the branching factror. However, when big clusters or molecular crystals are considered, $E_{bs}$ could be an unfeasible choice.

%\newpage

\section{The C$_{60}$-C$_{60}$H$_{28}$ complex}\label{app:big}

As for the water-methane dimer, single particle wavefunctions were obtained using a plane-wave cutoff of 300 Ry, and re-expanded in terms of B-splines with the natural grid spacing $a=\pi/G{\rm max}$. 
The Jastrow factor (e-e), (e-n) and (e-e-n) terms, and was constructed with the same procedure as in the water-methane system, i.e. by ensuring that it is the same in all systems. The variances of the VMC local energies were $\sim 11, 5.4$ and $5.8$ Ha$^2$ for the C$_{60}$-C$_{60}$H$_{28}$, C$_{60}$ and C$_{60}$H$_{28}$ systems, respectively.

\begin{figure}
\includegraphics[width=3.5in]{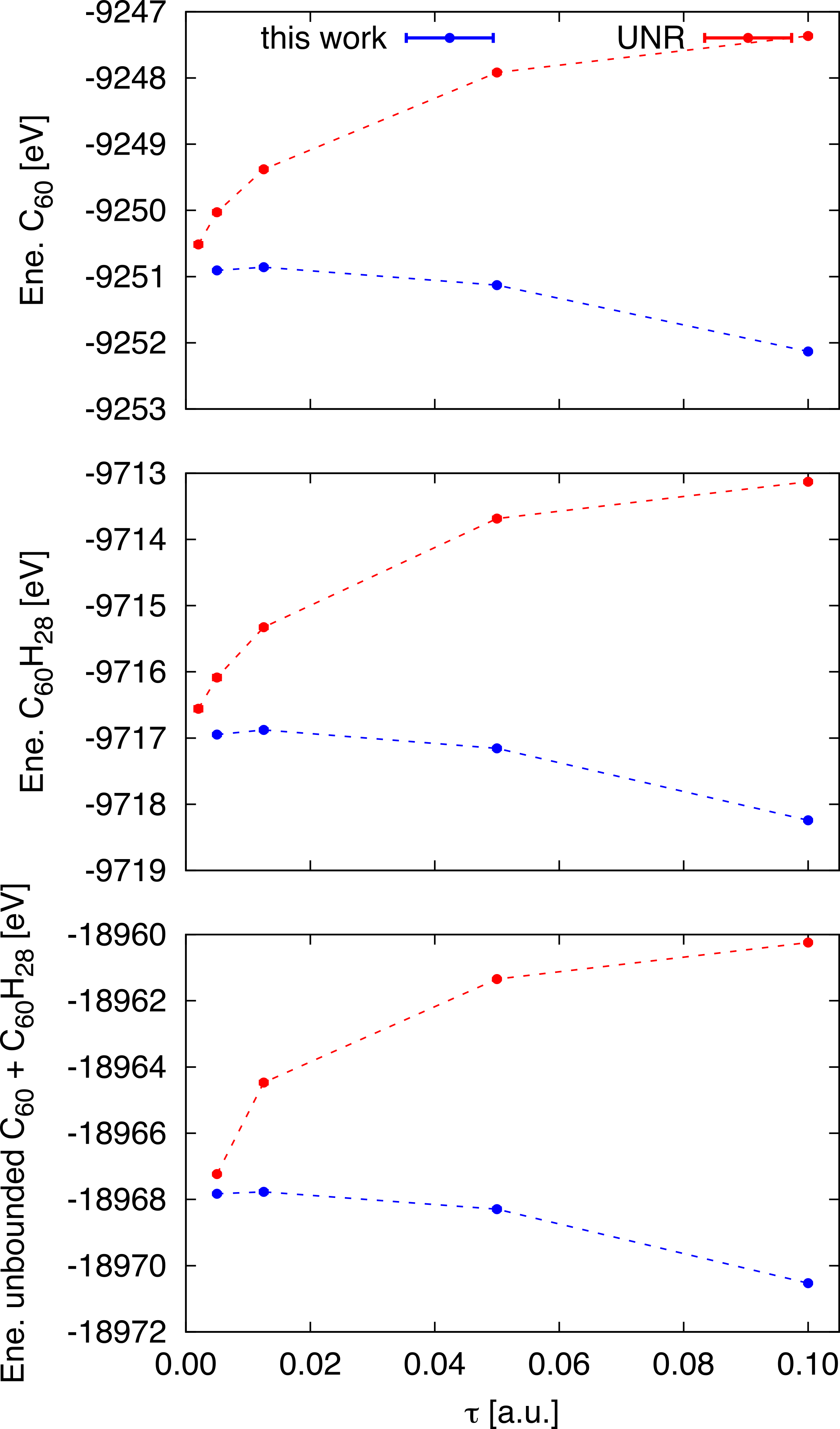}
\caption{\label{fig:Fig7} Energies of 
C$_{60}$ (top panel), C$_{60}$H$_{28}$(middle), and
unbonded C$_{60}$-C$_{60}$H$_{28}$ (bottom) systems as function of time step $\tau$, calculated using the UNR and present work prescriptions for the limits on the branching factor. Error bars are smaller than the size of the symbols.}
\end{figure}

In Fig.~\ref{fig:Fig7} we display the energy of the supramolecular system, $E_{\rm C_{60}-C_{60}H_{28}}$ as well as the energies of the monomers, $E_{\rm C_{60}}$ and $E_{\rm C_{60}H_{28}}$, computed in independent calculations performed with simulation cells containing either the isolated C$_{60}$ and C$_{60}$H$_{28}$ molecules, respectively. 

The improved accuracy of  present work prescriptions for the branching factor in comparison with the UNR branching factor can be appreciated in Fig.~\ref{fig:accuracy}(center).

%\newpage

\section{Two dimensional square ice}\label{app:ice}

We considered a monolayer of flat square ice of water, that is a system with 2-dimensional periodicity that is attaining considerable attention~\cite{Chen:2016gz,AlgaraSiller:2015cb}.
The unit cell include four water molecules, and here we considered a $4 \times 4$ supercell, for a total of 64 waters in the system.
The cohesive energy is obtained by subtracting the energy of the relevant number of isolated water molecules.
Single particle wavefunctions were obtained using a plane-wave cutoff of 600 Ry, and re-expanded in terms of B-splines with the natural grid spacing $a=\pi/G{\rm max}$. 
The larger plane-wave cutoff used for these calculations resulted in a lower variance of the VMC local energies, which was $\sim 0.28$~Ha$^2$ for the isolated molecule, and $\sim 19.8$~Ha$^2$ for the square ice (corresponding to $\sim 0.31$~Ha$^2$ per water molecule).
At the VMC level of theory the evaluated cohesive energy is -0.108(4)~eV, that is severely underestimated  (by a factor 4) with respect to the DMC evaluations.

\begin{figure}
\includegraphics[width=3.5in]{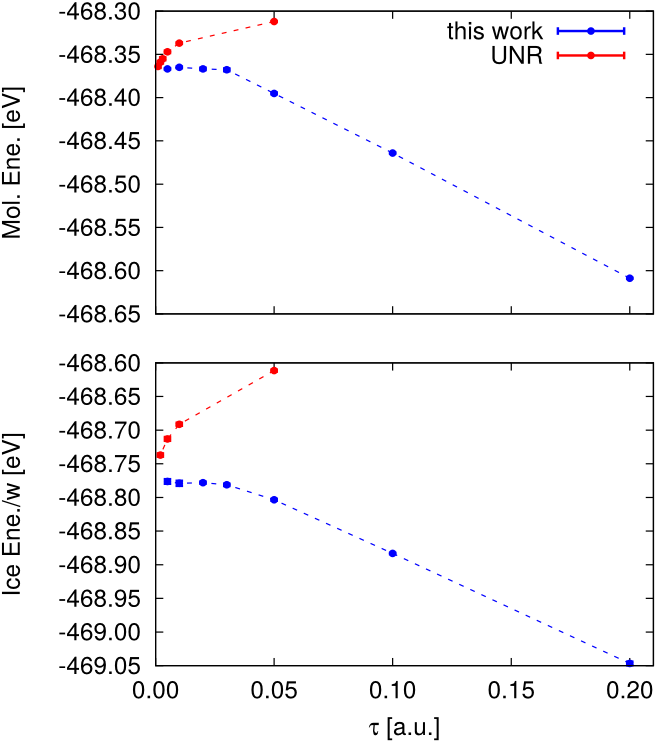}
\caption{\label{fig:ice} %2D-Ice 
Energies of an isolated water molecule (top), and of a water molecule in a periodic two dimensional squale ice (bottom) systems as function of time step $\tau$, calculated using the UNR and present work prescriptions for the limits on the branching factor. Error bars are smaller than the size of the symbols. }
\end{figure}

In Fig.~\ref{fig:ice} we display the energy of the isolated water molecule, as well as the energy per water in the square lattice 2-dimensional system. 
A comparison with Fig.~\ref{fig:Fig2} shows that the higher quality of the trial wavefunctions for this system results in a lower time step error.

The speedup obtained with  present work prescriptions for the branching factor in comparison with the UNR branching factor can be appreciated in Fig.~\ref{fig:accuracy}(left).

%\bibliography{ref}
\include{paper.bbl}

\end{document}